\documentclass[aps,prd,preprint,groupedaddress,showpacs]{revtex4}

\usepackage{epsfig}
\usepackage{graphics}
\usepackage{slashed}
\usepackage{color}
\usepackage[citecolor=true]{hyperref}
\usepackage{multirow}
\usepackage{amsmath}
\begin{document}

\leftmargin -2cm
\def\choosen{\atopwithdelims..}

\boldmath
\title{Production of three isolated photons in the parton Reggeization approach at high energies} \unboldmath

\author{\firstname{A.V.}\surname{Karpishkov}} \email{karpishkoff@gmail.com}
\author{\firstname{V.A.}\surname{Saleev}} \email{saleev@samsu.ru}

\affiliation{Samara National Research University, Moskovskoe Shosse,
34, 443086, Samara, Russia}

\affiliation{Joint Institute for Nuclear Research, Dubna, 141980
Russia}

%\author{\firstname{M.A.} \surname{Nefedov}} \email{nefedovma@gmail.com}
%\affiliation{Samara National Research University,
% Moskovskoe Shosse,
%34, 443086, Samara, Russia}

\begin{abstract}
We study a large-$p_T$ three-photon production  in proton-proton
collisions at the LHC. We use the leading order (LO) approximation
of the parton Reggeization approach consistently merged with the
next-to-leading order corrections originated  from the emission of
additional jet. For numerical calculations  we use the parton-level
generator KaTie and modified KMR-type unintegrated parton
distribution functions which satisfy exact normalization conditions
for arbitrary $x$. We compare our prediction with data from ATLAS
collaboration at the center-of-mass energy $\sqrt{s}=8$ TeV. We find
that the inclusion of the real next-to-leading-order corrections
leads to a good agreement between our predictions and data with the
same accuracy as for the next-to-next-to-leading calculations based
on the collinear parton model of QCD. At higher energies
($\sqrt{s}=13$ and 27 TeV) parton Reggeization approach predicts
larger cross sections, up to $\sim 15$ \% and $\sim 30$ \%,
respectively.
\end{abstract}

%Keywords: LHC, quantum chromodynamics, parton Reggeization approach,
%angular correlations, $\Upsilon$ meson, $D$ meson, single parton scattering, double parton scattering.\\

%\pacs{ 12.38.-t.}

\maketitle

\section{Introduction}
The recent experimental data for a large-$p_T$ multi-photon production
at the Tevatron \cite{CDF1,CDF2} and LHC \cite{LHC71,LHC72,LHC8} at
the energy range from 1.96 TeV up to 8 TeV  are extensively studied
in the collinear parton model (CPM) of perturbative quantum
chromodynamics (QCD)  beyond the leading-order (LO) accuracy in
strong-coupling constant, $\alpha_S$, i.e. at the
next-to-leading-order (NLO) \cite{NLO21,NLO31,NLO32} and even at
next-to-next-to-leading-order (NNLO)
\cite{NNLO21,NNLO22,NNLO31,NNLO32}. The high-order calculations for
two-photon or three-photon production in CPM of QCD provide rather
bad agreement with data at the level of NLO accuracy. For example,
NLO QCD calculations strongly underestimate, by factor 2 or even
more, recent data from ATLAS Collaboration at $\sqrt{s}=8$ TeV
\cite{LHC8} for three-photon production. The inclusion of additional
contribution from parton-shower mechanism and hadronization effects
\cite{NLO31} to the NLO calculations increase theoretical prediction
but they are being yet far from measured cross sections.

Inclusion of the NNLO QCD corrections for two-photon production
\cite{NNLO21} and three-photon production \cite{NNLO31,NNLO32}
eliminates the existing discrepancy with respect to NLO QCD
predictions. However, for three-photon production the agreement with
data is not so good as for two-photon production  and it is achieved
when hard scale parameter $\mu$ should be taken  very small
relatively usual used value \cite{NNLO31}.

In CPM of QCD we neglect the transverse momenta of initial-state
partons in hard-scattering amplitudes  that is a correct assumption
for the fully inclusive observables, such as $p_T$-spectra of single
prompt photons or jets, where their large transverse momentum
defines single hard scale of the process, $\mu \sim p_T$. The
corrections breaking the collinear factorization are shown to be
suppressed by powers of the hard scale \cite{Collins}.

The multi-photon large-$p_T$ production is multi-scale hard process
in which using the simple collinear picture of initial state
radiation may be a bad approximation. In the present paper, we
calculate different multi-scale variables in three-photon production
from a point of view of High-Energy Factorization (HEF) or
$k_T-$factorization, which initially has been introduced as a
resummation tool for $\ln (\sqrt{s}/\mu)$-enhanced corrections to
the hard-scattering coefficients in CPM, where invariants $\sqrt{s}$
referees to the total energy of process. We use the parton
Reggeization approach (PRA) which is a version of HEF formalism,
based on the  modified Multi-Regge Kinematics (mMRK) approximation
for QCD scattering amplitudes. This approximation is accurate both
in the collinear limit, which drives the
Transverse-Momentum-Dependent (TMD) factorization \cite{Collins} and
in the high-energy (Multi-Regge) limit, which is important for
Balitsky-Fadin-Kuraev-Lipatov(BFKL)~\cite{BFKL1,BFKL2,BFKL3,BFKL4}
resummation of $\ln (\sqrt{s}/\mu) $-enhanced effects.

In the same manner of the PRA, we have studied previously one-photon
production ~\cite{one-photon}, two-photon production
\cite{two-photon} and photon plus jet production ~\cite{photon-jet}
in proton-(anti)proton collisions at the Tevatron and the LHC. At
the present paper we study a production or three isolated photons at the
LHC. Preliminary, our predictions has been presented as short note
at DIS2021 Conference, see Ref. \cite{SaleevDIS}. The similar study
of three-photon production in the $k_T-$factorization approach was
published recently in Ref. \cite{three-photon}, where authors
compared predictions obtained with different unPDFs
\cite{KMR,WMR,PB1,PB2} that is compliment to our study in PRA
\cite{SaleevDIS}.

The paper has the following structure, in Section~\ref{sec:PRA} the
relevant basics of the PRA formalism are outlined. In the
Section~\ref{sec:KaTie} we overview Monte-Carlo (MC) parton-level
event generator KaTie and the relation between PRA and KaTie MC
calculations for tree-level amplitudes. In the
Section~\ref{sec:results} we compare obtained in the PRA results
with the recent ATLAS~\cite{LHC8} data as well as with theoretical
predictions obtained in NNLO calculations of CPM
\cite{NNLO31,NNLO32}. Our conclusions are summarized in the
Section~\ref{sec:conclusions}.

\section{Parton Reggeization Approach}

\label{sec:PRA}
%\subsection{High-energy factorization}
The PRA is based on high-energy factorization for hard processes in
the Multi-Regge kinematics. The basic ingredients of PRA are
$k_T-$dependent factorization formula, unintegrated parton
distribution functions (unPDF's) and gauge-invariant amplitudes with
off-shell initial-state partons.  The first one is proved in the
leading-logarithmic-approximation
 of high-energy QCD \cite{KTfactorization1,KTfactorization2}, the
second one is constructed in the same manner as it was suggested by
Kimber, Martin, Ryskin and Watt \cite{KMR,WMR}, but with sufficient
revision \cite{NefedovSaleev2020}. The off-shell amplitudes are
derived using the Lipatov's Effective Field Theory (EFT) of
Reggeized gluons~\cite{Lipatov95} and Reggeized
quarks~\cite{LipatovVyazovsky}. The brief description of LO in
$\alpha_S$ approximation of PRA is presented below. More details can
be found in Refs.~\cite{NSS2013,NKS2017}, the inclusion of real NLO
corrections in the PRA was studied in Ref.~\cite{NKS2017}, the
development of PRA in the full one-loop NLO approximation was
further discussed in \cite{PRANLO1,PRANLO2,PRANLO3}.

Factorization formula of PRA in LO approximation for the process
$p+p\to {\cal Y}+ X$, can be obtained from factorization formula of
the CPM for the auxiliary hard subprocess like $g+g\to q + {\cal Y}+
\bar q$. For discussed here process of three-photon production,
${\cal Y}=\gamma\gamma\gamma$. In the Ref.~\cite{NKS2017} the
modified Multi-Regge Kinematics (mMRK) approximation for the
auxiliary amplitude has been constructed, which correctly reproduces
the Multi-Regge and collinear limits of corresponding QCD amplitude.
This mMRK-amplitude has $t$-channel factorized form, which allows
one to rewrite the cross-section of auxiliary subprocess in a
$k_T$-factorized form:
  \begin{eqnarray}
  d\sigma &=& \sum_{i,\bar j}\int\limits_0^1 \frac{dx_1}{x_1} \int \frac{d^2{\bf q}_{T1}}{\pi} \tilde{\Phi}_i(x_1,t_1,\mu^2)
\int\limits_0^1 \frac{dx_2}{x_2} \int \frac{d^2{\bf q}_{T2}}{\pi}
\tilde{\Phi}_{j}(x_2,t_2,\mu^2)\cdot d\hat{\sigma}_{\rm PRA},
\label{eqI:kT_fact}
  \end{eqnarray}
where $t_{1,2}=-{\bf q}_{T1,2}^2$, the off-shell partonic
cross-section $\hat\sigma_{\rm PRA}$ in PRA is determined by squared
PRA amplitude, $\overline{|{\cal A}_{PRA}|^2}$. Despite the fact
that four-momenta ($q_{1,2}$) of partons in the initial state of
amplitude  ${\cal A}_{PRA}$ are off-shell ($q_{1,2}^2=-t_{1,2}<0$),
the PRA hard-scattering amplitude is gauge-invariant because the
initial-state off-shell partons are treated as Reggeized partons of
gauge-invariant EFT for QCD processes in Multi-Regge Kinematics
(MRK), introduced by L.N. Lipatov
in~\cite{Lipatov95,LipatovVyazovsky}. The Feynman rules of this EFT
are written down in Refs.~\cite{LipatovVyazovsky,AntonovLipatov}.

%\subsection{Modified KMR unintegrated PDFs in multi-Regge kinematics}

 The tree-level unPDFs~ $\tilde{\Phi}_{i}(x_{1,2},t_{1,2},\mu^2)$ in Eq. (\ref{eqI:kT_fact})
 are equal to the convolution
of the collinear PDFs $f_{i}(x,\mu^2)$ and
Dokshitzer-Gribov-Lipatov-Altarelli-Parisi (DGLAP) splitting
function $P_{ij}(z)$ with the factor $1/t_{1,2}$,
\begin{equation}
\tilde{\Phi}_i(x,t,\mu)= \frac{\alpha_s(\mu)}{2\pi} \frac{1}{t}
\sum\limits_{j=q,\bar{q},g}\int\limits_x^1 dz\ P_{ij}(z) {F}_j\left(
\frac{x}{z}, \mu_F^2 \right),
%\theta\left(
%\Delta(t,\mu_Y^2)-z \right),
\end{equation}
where $F_i(x,\mu_F^2)=x f_j(x,\mu_F^2)$. Here and above we put
$\mu_F=\mu_R=\mu$. Consequently, the cross-section
(\ref{eqI:kT_fact}) with such unPDFs contains the collinear
divergence at $t_{1,2}\to 0$ and infrared (IR) divergence at
$z_{1,2}\to 1$.

To resolve collinear divergence problem of $\tilde{\Phi}_i(x,t,\mu)$
we require that modified unPDF ${\Phi}_i(x,t,\mu)$ should be
satisfied exact normalization condition:
\begin{equation}
\int\limits_0^{\mu^2} dt \Phi_i(x,t,\mu^2) = {F}_i(x,\mu^2),
\end{equation}
which is equivalent to:
\begin{equation}
\Phi_i(x,t,\mu^2)=\frac{d}{dt}\left[ T_i(t,\mu^2,x){F}_i(x,t)
\right],\label{eq:sudakov}
\end{equation}
where $T_i(t,\mu^2,x)$  is usually referred to as Sudakov
form-factor, satisfying the boundary conditions $T_i(t=0,\mu^2,x)=0$
and $T_i(t=\mu^2,\mu^2,x)=1$. Such a way, modified unPDF can be
written as follows from KMR model:
\begin{equation}
\Phi_i(x,t,\mu)= \frac{\alpha_s(\mu)}{2\pi} \frac{T_i(t,\mu^2,x)}{t}
\sum\limits_{j=q,\bar{q},g}\int\limits_x^1 dz\ P_{ij}(z) {F}_j\left(
\frac{x}{z}, t \right) \theta\left( \Delta(t,\mu)-z
\right).\label{uPDF}
\end{equation}
Here, we resolved also IR divergence taking into account observation
that the mMRK expression gives a reasonable approximation for the
exact matrix element only in the rapidity-ordered part of the
phase-space. From this requirement, the following cutoff on
$z_{1,2}$ can be derived: $z_{1,2}< 1-\Delta_{KMR}(t_{1,2},\mu^2),$
where $\Delta_{KMR}(t,\mu^2)=\sqrt{t}/(\sqrt{\mu^2}+\sqrt{t})$ is
the KMR-cutoff function~\cite{KMR}.

The solution for Sudakov form-factor in Eq. ({\ref{eq:sudakov}) has
been obtained in Ref.~\cite{NefedovSaleev2020}:
\begin{equation}
T_i(t,\mu^2,x)=\exp\left[ -\int\limits_t^{\mu^2} \frac{dt'}{t'}
\frac{\alpha_s(t')}{2\pi} \left( \tau_i(t',\mu^2) + \Delta\tau_i
(t',\mu^2,x) \right) \right]\label{eq:sud}
\end{equation}
with
\begin{eqnarray*}
\tau_i(t,\mu^2)&=&\sum\limits_j \int\limits_0^1 dz\ zP_{ji}(z)\theta (\Delta(t,\mu^2)-z), \label{eq:tau} \\
\Delta\tau_i(t,\mu^2,x)&=& \sum\limits_j \int\limits_0^1 dz\
\theta(z-\Delta(t,\mu^2)) \left[ zP_{ji}(z) -
\frac{{F}_j\left(\frac{x}{z},t \right)}{{F}_i(x,t)} P_{ij}(z)
\theta(z-x) \right].
\end{eqnarray*}
Let us summarize important differences between the Sudakov
form-factor obtained in our mMRK approach (\ref{eq:sud}) and in the
KMR approach~ \cite{KMR}. At first, the Sudakov form-factor
(\ref{eq:sud})  contains the $x-$depended $\Delta \tau_i$-term in
the exponent which is needed to preserve exact normalization
condition for arbitrary $x$ and $\mu$. The second one is a
numerically-important difference that in our mMRK approach the
rapidity-ordering condition is imposed both on quarks and gluons,
while in KMR approach it is imposed only on gluons.

To illustrate differences between unPDFs at large $x$, obtained in
original KMR~\cite{KMR,WMR} model and in our modified approach
   \cite{NefedovSaleev2020}, we plot ratios for integrated over transverse momentum unPDFs to parent collinear PDFs for
gluon and $u-$quark as function of $x$ at different choice of hard
scale $\mu$ in Figures \ref{fig:PDF1} and \ref{fig:PDF2},
correspondingly.

In contrast to most of studies in the $k_T$-factorization, the
gauge-invariant matrix elements with off-shell initial-state partons
(Reggeized quarks and gluons) from Lipatov's EFT~\cite{Lipatov95,
LipatovVyazovsky} allow one to study arbitrary processes involving
non-Abelian structure of QCD without violation of Slavnov-Taylor
identities due to the nonzero virtuality of initial-state partons.
This approach, together with KMR-type unPDFs gives stable and
consistent results in a wide range of phenomenological applications,
which include the description of the angular correlations of
dijets~\cite{NSS2013},
charmed~\cite{Maciula:2016wci,Karpishkov:2014epa} and
bottom-flavored~\cite{NKS2017,Karpishkov:2016hnx} mesons, charmonia
\cite{Saleev:2012hi,He:2019qqr} as well as some other examples.

\section{Details of numerical calculations}
\label{sec:KaTie}

The first step of calculations in PRA is generation of amplitudes of
relevant off-mass shell partonic processes by Feynman rules of
Lipatov's EFT. It can be done using a model file
ReggeQCD~\cite{ReggeQCD} for FeynArts tool~\cite{hahn}. In the Fig.~\ref{fig:3gamma1}, the total set of
 13 Feynman diagrams for LO process
\begin{equation}
Q\bar Q \to \gamma\gamma\gamma \label{QQ3gamma}
\end{equation}
obtained with ReggeQCD is shown. The number of EFT diagrams for NLO in $\alpha_S$ involved
in our study process
\begin{equation}
Q R\to q\gamma\gamma\gamma \label{QR3gammaQ}
\end{equation}
 is getting too large
for analytical calculation.  In Fig.~\ref{fig:3gamma2}, the full
gauge invariant set of 40 Feynman diagrams is shown. To proceed next
step, we should analytically calculate squared off-shell amplitudes
and perform a numerical calculation using factorization
formula~(\ref{eqI:kT_fact}) with modified unPDFs~(\ref{uPDF}). At
present, we can do it with  required numerical accuracy only for
$2\to 2$ and $2 \to 3$ off-shell parton processes. To calculate
contributions from $2 \to 4$ processes with initial Reggeized
partons we use parton-level generator KaTie~\cite{katie}.

A few years ago, a new approach to derive gauge-invariant
scattering amplitudes with off-shell initial-state partons for
high-energy scattering, using the spinor-helicity techniques and
BCFW-like recursion relations for such amplitudes has been
introduced in the Refs.~\cite{hameren1,hameren2}. Some time later
the MC parton-level event generator KaTie \cite{katie} has been
developed to provide calculations for hadron scattering processes
that can deal with partonic initial-state momenta with an explicit
transverse momentum dependence causing them to be space-like. The
formalism ~\cite{hameren1,hameren2} for numerical generation of
off-shell amplitudes is equivalent to the results of Lipatov's EFT at
the tree level \cite{NSS2013,NKS2017,kutak}.  We should note here,
that for  the generalization of the formalism to full NLO
level~\cite{PRANLO1, PRANLO2}, the use of explicit Feynman rules and
the structure of EFT is more convenient.

Taking in mind above mentioned discussion, the LO contribution of
subprocess (\ref{QQ3gamma})  has been calculated  for crosscheck
both with KaTie MC generator and using direct integration of
analytical squared amplitudes obtained with the help of Feynman
rules of Lipatov EFT. All final calculations have been done using MC
event generator KaTie \cite{katie}.

We will neglect NLO contribution in quark-antiquark annihilation
channel from subprocesses with additional final gluon
\begin{equation}
Q\bar Q\to g \gamma\gamma\gamma \label{QQ3gammaG},\end{equation}
 which should be negligibly
small in comparison with main others as in the similar case of NLO
CPM calculations. First off all, because the relevant values of
involving longitudinal parton momenta are very small ($x<10^{-2}$)
at the energy range of the LHC and the gluon density is much larger
 than the quark (antiquark) ones. Such a way, we avoid
difficulties in a calculation of the process (\ref{QQ3gammaG}),
which follow from an infra-red divergence, which should be
regularized by a contribution from loop correction to the LO process
(\ref{QQ3gamma}) and from double counting between LO
(\ref{QQ3gamma}) and NLO (\ref{QQ3gammaG}) diagrams with emission of
an additional gluon. The technique of NLO calculations is still
under development in PRA, see discussions in
Refs.~\cite{PRANLO1,PRANLO2, PRANLO3}.

 The next important issue is that a calculation for the process
(\ref{QR3gammaQ}) doesn't contain infra-red and collinear
singularities in PRA, after taking in consideration isolation-cone
conditions for final photons and partons. Numerical accuracy of
total cross section calculations with MC generator KaTie by default
is 0.1 \% .

\section{Results}
\label{sec:results}

First of all, we review setup of ATLAS measurements at $\sqrt{s}=8$
 TeV \cite{LHC8}:
\begin{itemize}
\item Photon transverse energies  (transverse momenta) (1 is leading photon, 2 is sub-leading photon and 3 is sub-sub-leading photon)
  $E_{T1}>27$ GeV, $E_{T2}>22$ GeV, $E_{T3}>15$
GeV.
\item  For rapidity (pseudorapidity) of all photons one has $|\eta_{1,2,3}|<2.37$, excluding the range
$1.37<|\eta_{1,2,3}|<1.56$.
\item Three-photon invariant mass $M_{123}=M_{3\gamma}>50$~GeV.
\item Photon-photon isolation conditions are $\Delta R_{ij}>R_{\gamma\gamma}=0.45$, where $\Delta R_{ij}=\sqrt{(\eta_i-\eta_j)^2+(\phi_i-\phi_j)^2}$
\item Photon-quark isolation conditions are $\Delta R_{iq}>R_0=0.40$
\end{itemize}
To take into account a fragmentation contribution, we use the Frixione
smooth photon isolation~\cite{Frixione}. For any angular difference
$\Delta R_{iq}$ from each photon, when $\Delta R_{iq}\leq R_0$, it
is required
$$E_T^{iso}(\Delta R_{iq})<E_T^{max}\frac{1-\cos(\Delta R{iq})}{1-\cos (R_0)},$$
where $E_T^{max}=10$~GeV, $E_T^{iso}=E_{Tq}$.

We test dependence of predicted cross section on
choice of factorization ($\mu_F$)and renormalization ($\mu_R$)
scales, which we take equal to each other, $\mu_F=\mu_R=\mu$. In the
Table~\ref{Tab:1} we compare predictions obtained with
$\mu=M_{3\gamma}$ -- an invariant mass of the three-photon system and
$\mu=E_{T,\sum}=E_{T,1\gamma}+E_{T,2\gamma}+E_{T,3\gamma}$ -- a sum of
transverse momenta (transverse energies) moduli of photons. Errors
indicate upper and lower limits of the cross section obtained due to
variation of the hard scale $\mu$ by a factor $\xi=2$ or $\xi=1/2$ around the
central value of the hard scale.

% and
%$\mu=k_{T,1\gamma}$ - highest transverse momentum of photons.

\begin{table} \label{Tab:1}
\begin{tabular}{|c|c|c|c|} \hline
Hard scale, $\mu$ & $\sigma_{LO}(Q\bar Q\to 3\gamma)$ [fb]  &
$\sigma(QR\to 3\gamma q)$ [fb] & $\sigma_{NLO}$ [fb]
\\ \hline

 $M_{3\gamma}$ & $37.20^{+9.25}_{-7.98} $ &
 $36.94^{-6.14}_{+5.91}$ & $73.14^{+4.13}_{-1.07}$\\ \hline

$E_{T,\sum}$ & $36.35^{+8.38}_{-9.77}$   & $39.26^{-6.29}_{+6.00}$ &
$75.62^{+3.59}_{-2.39}$
\\ \hline

% $E_{T,3\gamma}/2$  &
%$32.50^{+9.80}_{-2.65}$  & $71.00^{+4.93}_{-2.65}$  \\ \hline
\end{tabular}
\caption{PRA predictions for $p+p\to \gamma\gamma\gamma + X$ total
cross section at $\sqrt{s}=8$~TeV for the different choice of
factorization/renormalization scale ($\mu=\mu_F=\mu_R$). Errors
indicate upper and lower limits of the cross section due to scale
uncertainty.}
\end{table}

\begin{table} \label{Tab:2}
\begin{tabular}{|c|c|c|c|c|} \hline
$\sqrt{s}$[TeV] & $\sigma_{LO}(Q\bar Q\to 3\gamma)$ [fb]  &
$\sigma(QR\to 3\gamma q)$ [fb] & $\sigma_{NLO}$ [fb] &
$\sigma_{\mbox{\small NNLO}}^{\mbox{\small CPM}}$ \cite{NNLO32}
\\ \hline
8  &  $37.20^{+9.25}_{-7.98} $ &
 $36.94^{-6.14}_{+5.91}$ & $73.14^{+4.13}_{-1.07}$ & $67.42^{+7.41}_{-5.73}$\\
\hline
 13 & $61.64^{+16.88}_{-15.63}$ & $72.87^{-9.72}_{+10.78}$  & $134.51^{+6.10}_{-3.91}$  & $114^{+13.64}_{-10.54}$\\ \hline
27 &  $132.03^{+40.52}_{-35.50}$ & $192.96^{-24.61}_{+19.07}$   & $324.99^{+15.91}_{-16.43}$  & $245.91^{+32.46}_{-24.34}$\\
\hline
\end{tabular}
\caption{Predictions for $p+p\to \gamma\gamma\gamma + X$ total cross
section at the different center-of-mass energies, $\sqrt{s}$. Hard
scale is taken as $\mu=M_{3\gamma}$. Numerical error of total cross
section calculation is equal to $0.1\%$.}
\end{table}

As we see in Table \ref{Tab:1}, where the total cross sections of
three-photon production are presented, relative contribution of LO
subprocesses grows with increase of the hard scale $\mu$ and
contribution of NLO subprocesses oppositely falls down, however their
sum changes only a little. Predicted absolute values of
cross-section are in a quite well agreement with the experimental data~\cite{LHC8}
as well as with the NNLO CPM results \cite{NNLO31,NNLO32}
taking in mind the level of accuracy, which is originated from scale
variation.

At higher energies, $\sqrt{s}=13$~TeV and $\sqrt{s}=27$~TeV, the PRA
predicts larger cross sections in comparing with the NNLO CPM
calculations, see Table~\ref{Tab:2}. We estimate excess
approximately in 15 and 30 \%, correspondingly.  In the PRA we
obtain also a strong decreasing of scale uncertainty in the NLO
approximation instead of the LO one as it is estimated from general
properties of perturbative QCD. In fact, one has LO scale
uncertainty is about 25-30 \%, but at NLO level of calculation it is
only 4-5 \% at different energies. Let us note that in NNLO CPM
calculation of three-photon production \cite{NNLO31,NNLO32} such
uncertainty is still about 10 \%.

Differential spectra, which demonstrate different kinematics
correlations between final photons, are shown in Figures
(\ref{fig:M123}) - (\ref{fig:Y12Y13}). There are no kinematics
regions in invariant masses, pseudo-rapidities, azimuthal angles or
transverse momenta where one of the relevant contributions can been
considered as an absolutely dominant one. To describe the data, only
both should be taken.
 The NLO  contribution in $\alpha_s$~(\ref{QR3gammaQ}) is
enhanced evidently because it is proportional to a quark-gluon
luminosity instead of a quark-antiquark luminosity in case of LO
production~(\ref{QQ3gamma}) in proton-proton collision.

\section{Conclusions}
\label{sec:conclusions}

We obtain a quite satisfactory description for cross section and
spectra for the three-photon production in the LO PRA with a matching of a real
NLO correction from partonic subprocess (\ref{QR3gammaQ}) at the
$\sqrt{s}=8$~TeV. We demonstrate an applicability of the new KMR-type quark
and gluon unPDFs to use in high-energy factorization calculations. It
has been shown that, as in our previous studies of hard processes in the
PRA, obtained results in LO approximation coincide with full NLO
predictions of the CPM and, respectively, NLO calculations in the PRA
roughly reproduce NNLO predictions of the CPM. However, at higher energies
(13 and 27~TeV) the PRA predicts larger cross sections, up to $\sim 15$
\% and $\sim 30$ \%, with respect to predictions of the NNLO CPM. The last
fact can be used for a discrimination between the high-energy
factorization and the collinear factorization for hard processes at high
energies.

\section*{Acknowledgments}
We are grateful to A. van Hameren  for helpful communication on
 MC generator KaTie, M. Nefedov and A. Shipilova for useful physics discussions.
The work has been supported in parts by the Ministry of Science and Higher Education of Russia via State assignment to
educational and research institutions under project FSSS-2020-0014.

%\end{document}

\clearpage
\begin{figure}[h!]
\begin{center}
\includegraphics[width=0.5\textwidth]{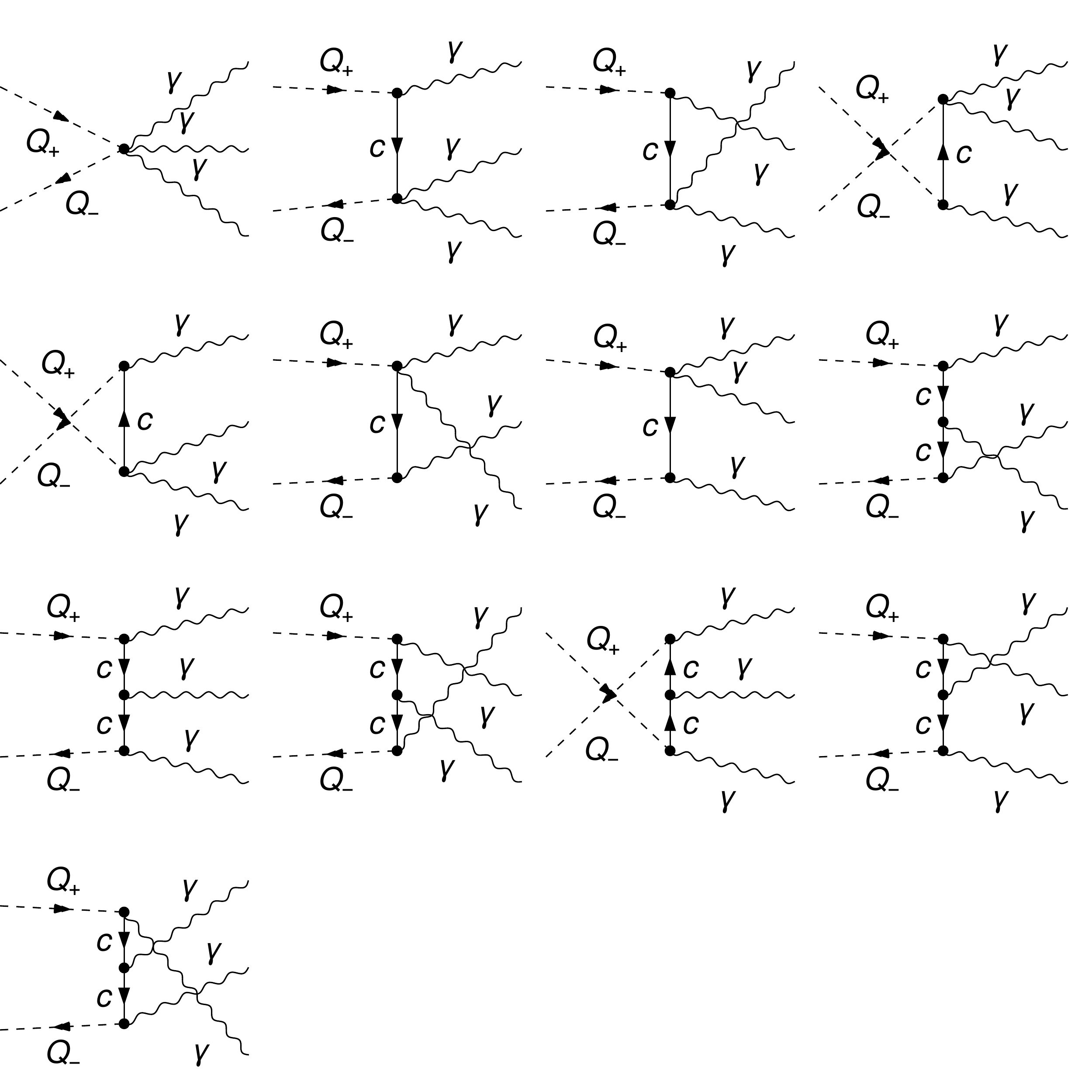}
\end{center}
\caption{The total set of 13 Feynman diagrams for $Q\bar Q \to
\gamma\gamma\gamma$ obtained with ReggeQCD~\cite{ReggeQCD}.
\label{fig:3gamma1}}
\end{figure}

\begin{figure}[h!]
\begin{center}
\includegraphics[width=\textwidth,angle=0,trim= 0cm 0cm 0cm 3cm,clip]{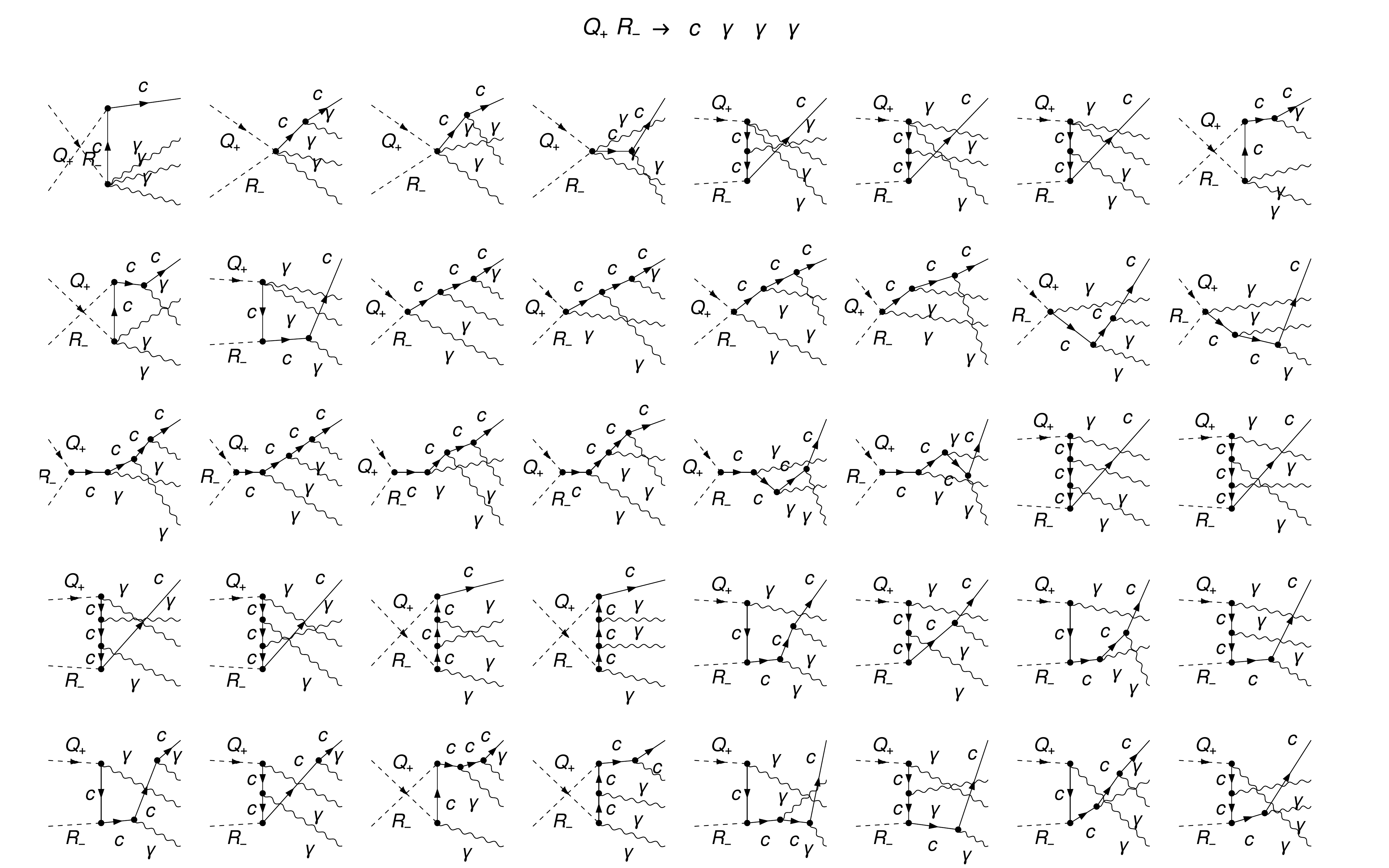}
\end{center}
\caption{The total set of 40 Feynman diagrams for $Q R\to
q\gamma\gamma\gamma$ obtained with
ReggeQCD~\cite{ReggeQCD}.\label{fig:3gamma2}}
\end{figure}

\begin{figure}[h!]
  \begin{center}
\includegraphics[width=0.6\textwidth]{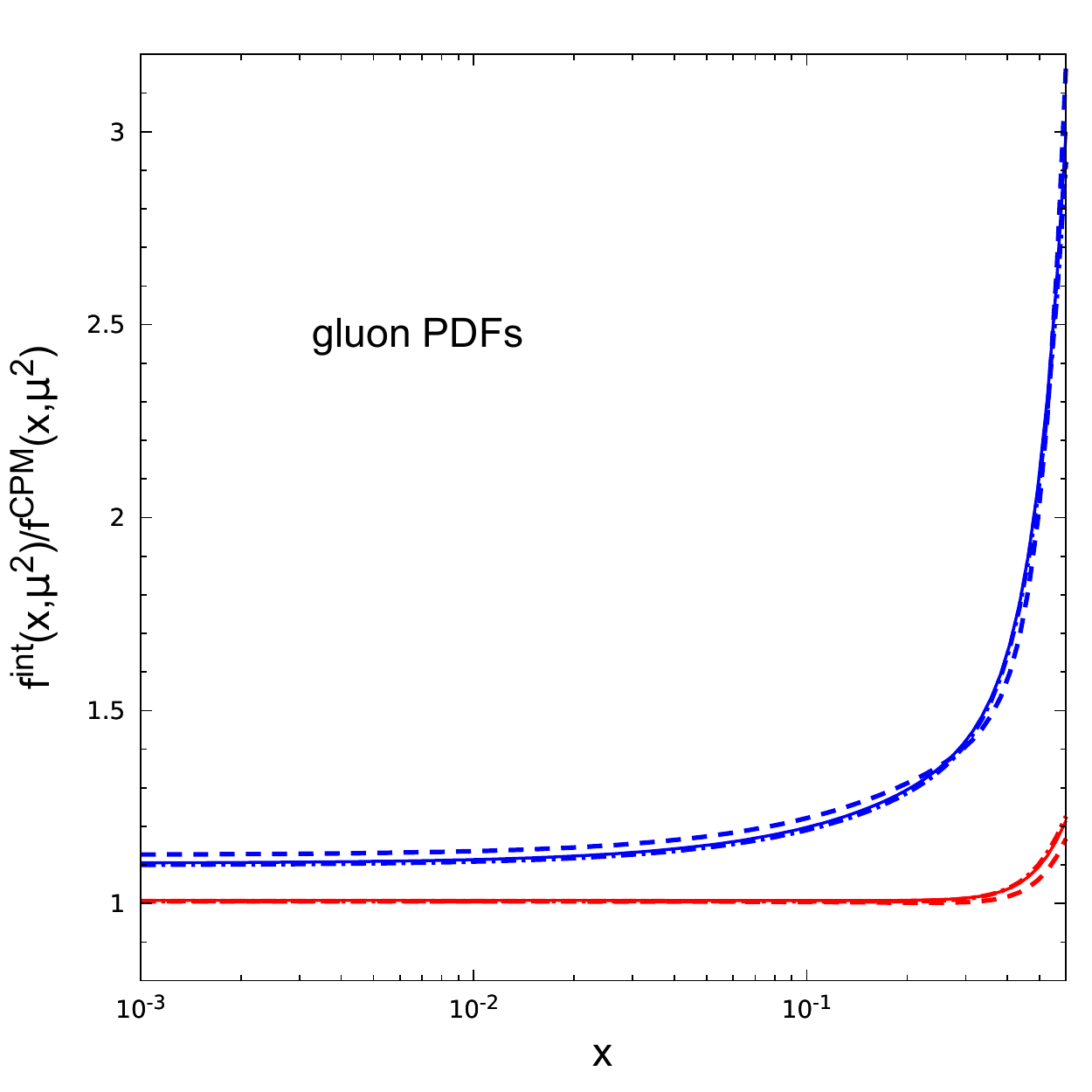}
   \end{center}
  \caption{Ratio for integrated over transverse momentum unPDF to parent collinear PDF for
gluon as function of $x$ at different choice of hard scale
$\mu^2=10^4, 6\times 10^4, 10^5$  GeV$^2$ which correspond to
dashed, solid and dotted-dashed lines. Blue lines are obtained in
original KMR  model \cite{KMR,WMR} and red lines are obtained in our
modified approach \cite{NefedovSaleev2020}.
    \label{fig:PDF1}}
  \end{figure}

\begin{figure}[h!]
  \begin{center}
 \includegraphics[width=0.6\textwidth]{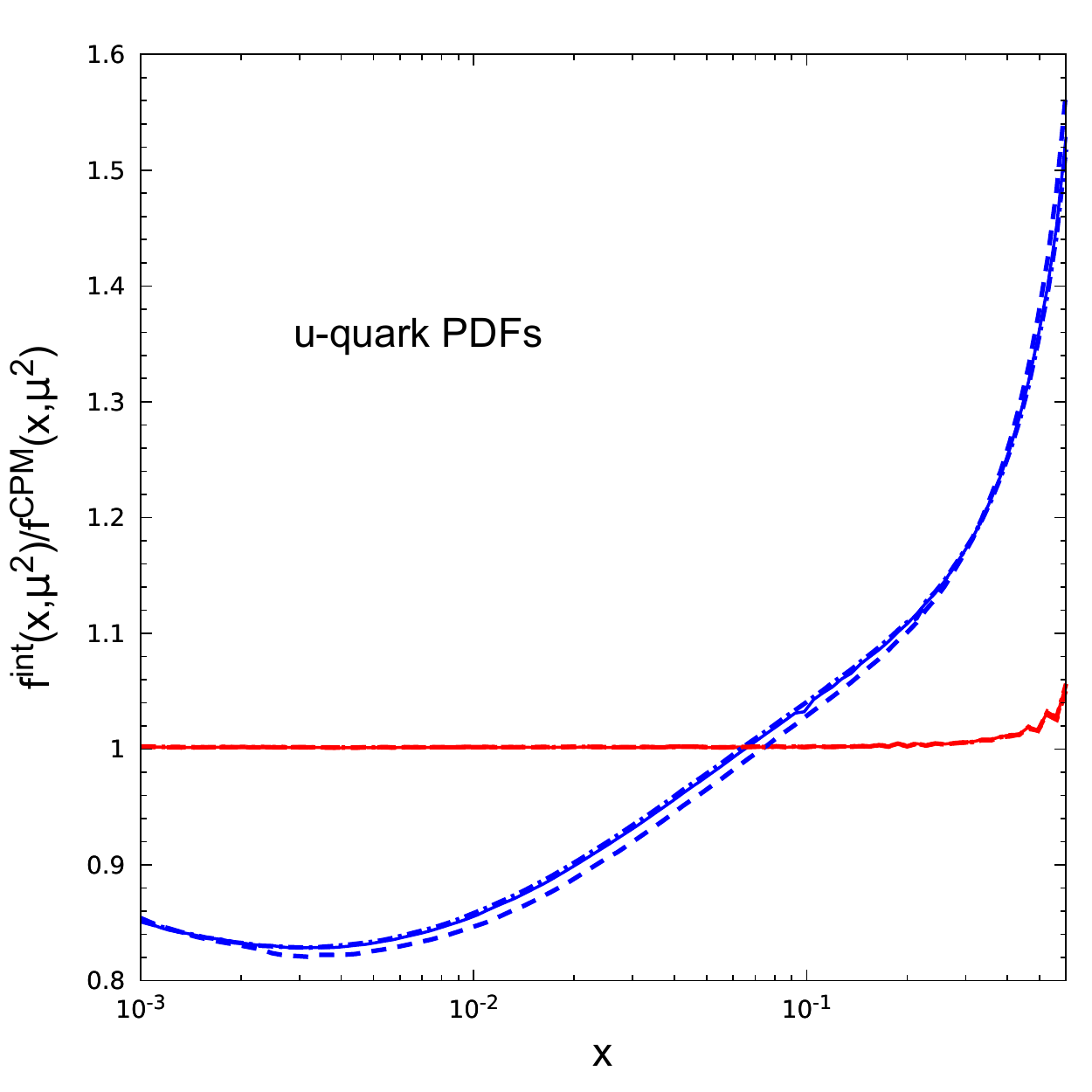}
   \end{center}
  \caption{Ratio for integrated over transverse momentum unPDF to parent collinear PDF
  for valence
$u-$quark as function of $x$ at different choice of hard scale
$\mu^2=10^4, 6\times 10^4, 10^5$  GeV$^2$, which correspond to
dashed, solid and dotted-dashed lines. Blue lines are obtained in
original KMR  model \cite{KMR,WMR} and red lines are obtained in our
modified approach \cite{NefedovSaleev2020}.
    \label{fig:PDF2}}
  \end{figure}

\begin{figure}[p!]
  \begin{center}
      \includegraphics[width=0.45\textwidth]{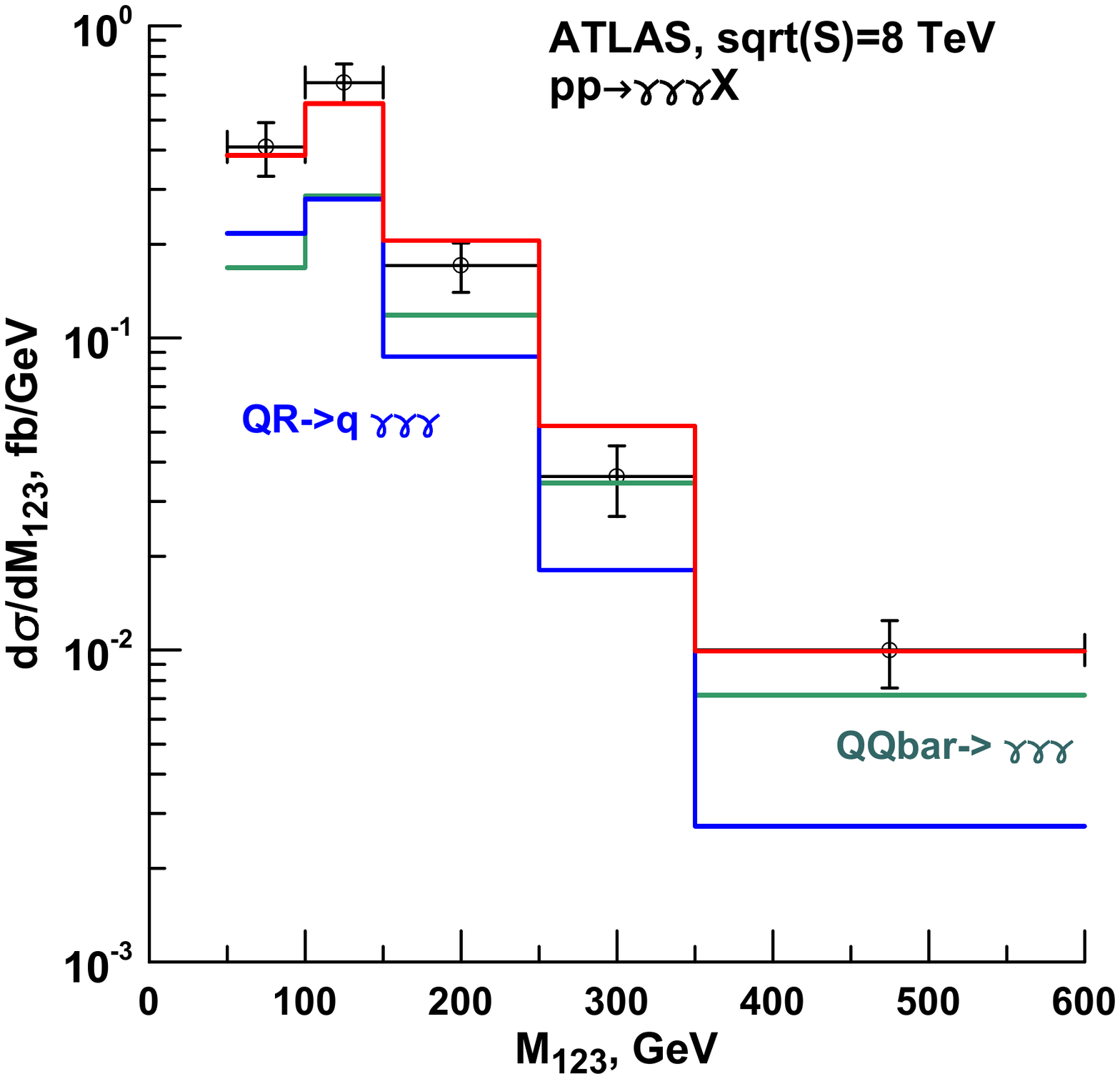}
  \end{center}
  \caption{The differential cross sections for the production of three isolated photons as functions
   of $M_{123}=M_{3\gamma}$.
   The hard scale in PRA calculations are taken as invariant mass of three-photon  system,
   $\mu_0=M_{3\gamma}$.
   The green histogram corresponds LO contribution from $Q\bar Q \to \gamma\gamma\gamma$ subprocess. The blue histogram corresponds NLO
   contribution from $QR\to q\gamma\gamma\gamma$ subprocess. The
   red histogram is their sum..
    \label{fig:M123}}
  \end{figure}

\clearpage

 \begin{figure}[p!]
  \begin{center}
  \includegraphics[width=0.3\textwidth]{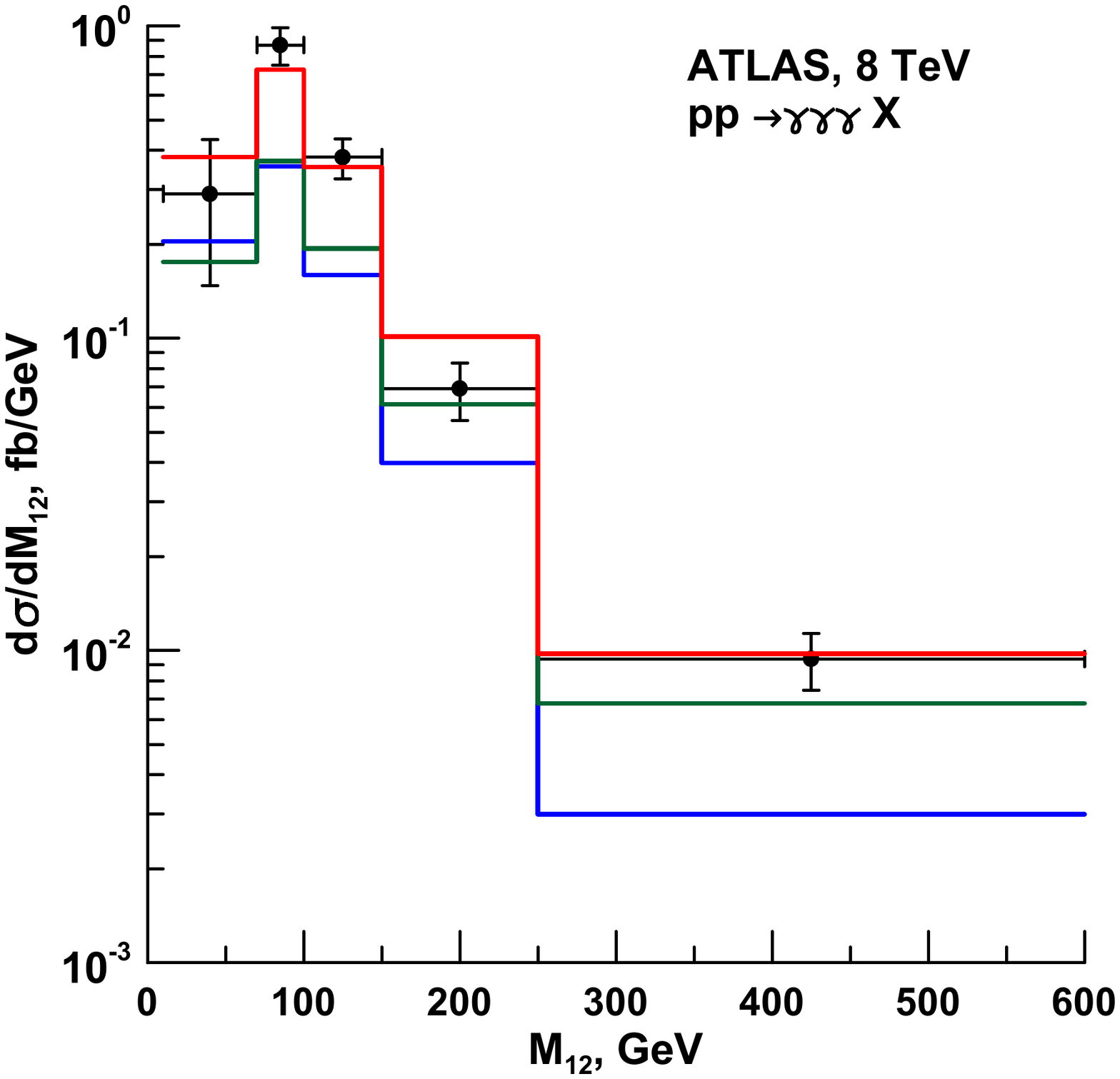}
  \includegraphics[width=0.3\textwidth]{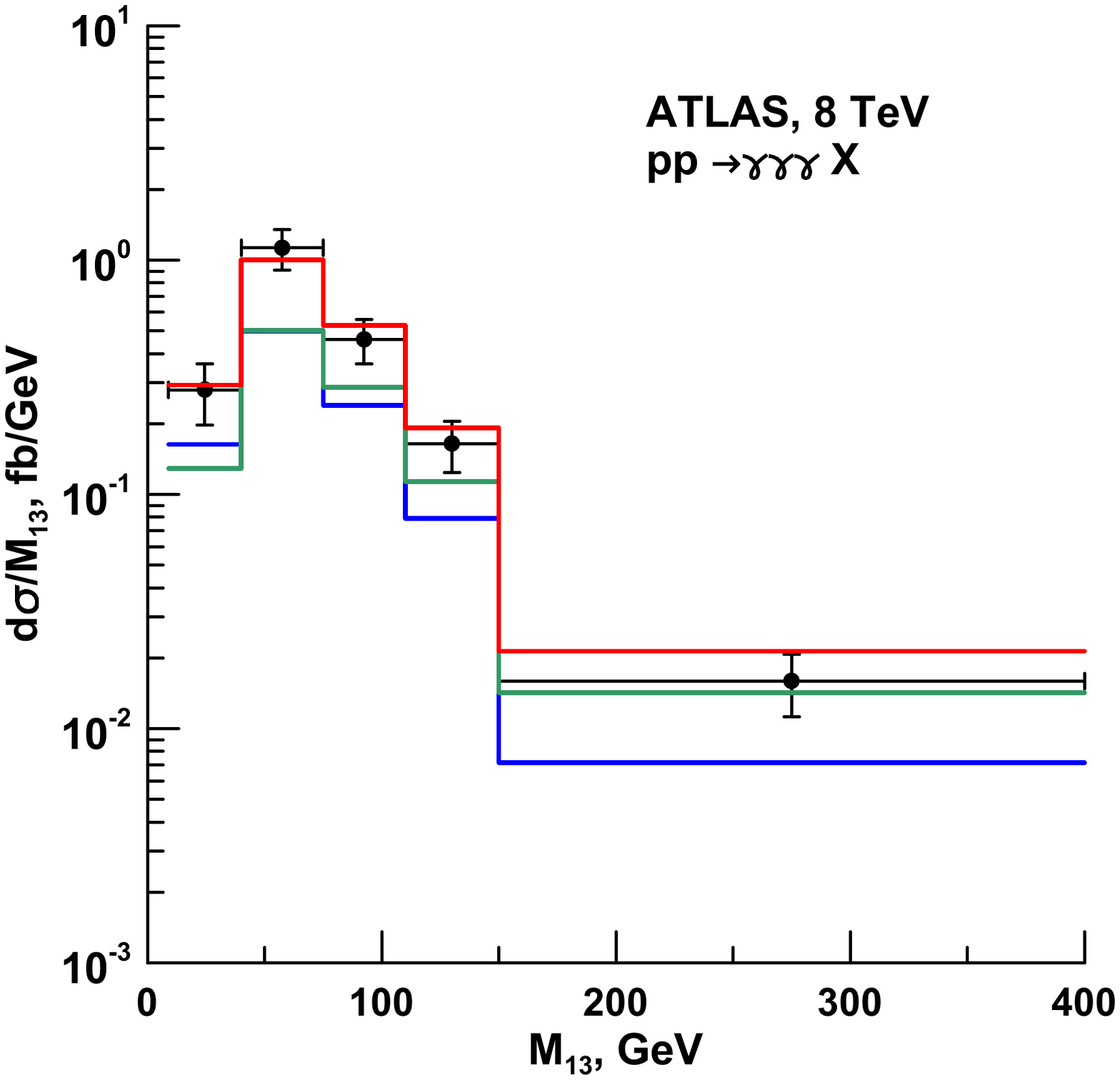}
  \includegraphics[width=0.3\textwidth]{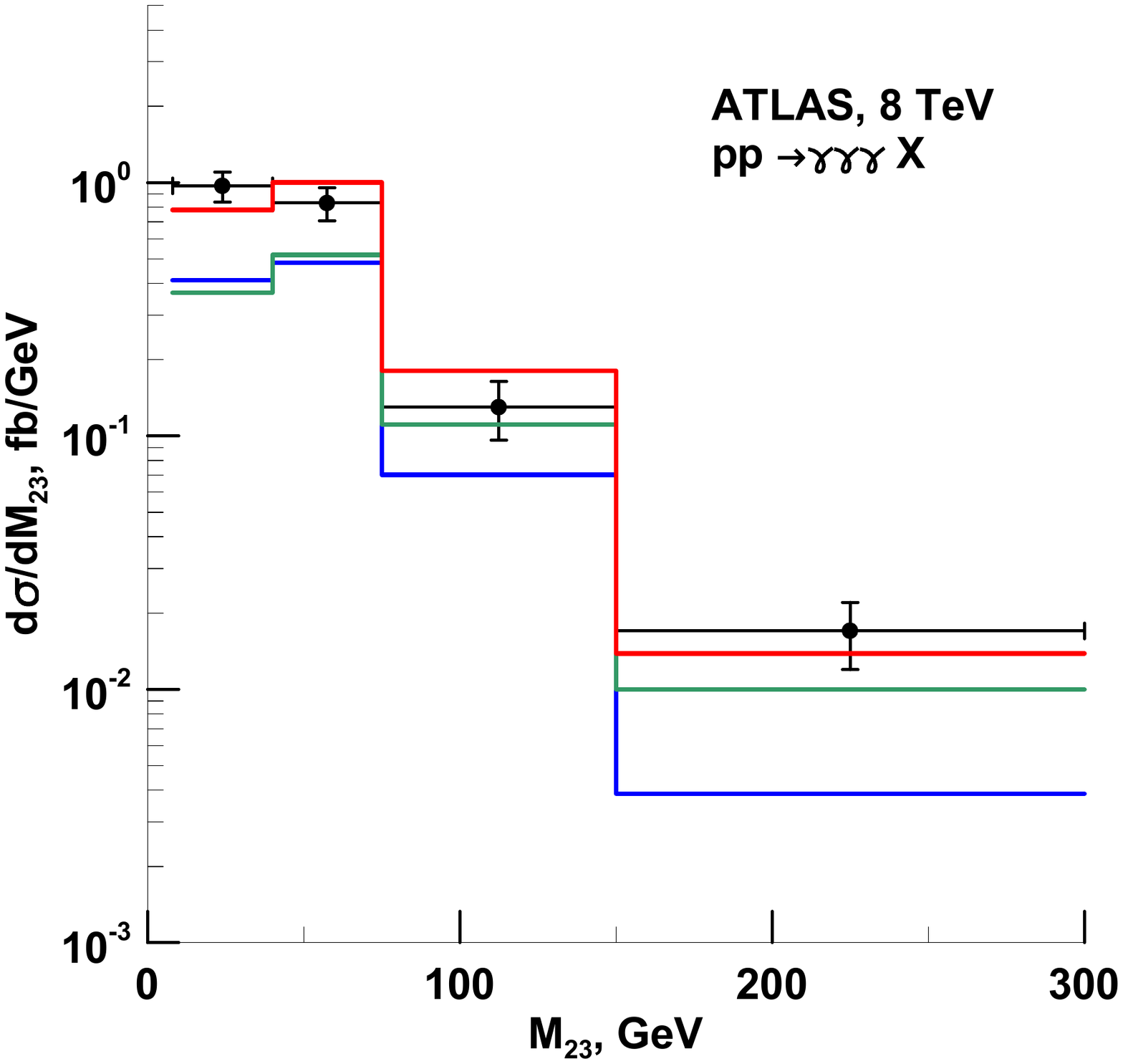}
      \end{center}
  \caption{The differential cross sections for the production of three isolated photons as functions
   of $M_{12}$ (left panel), $ M_{13}$ (central panel), $M_{23}$ (right panel).
  Curves are defined as in Fig. \ref{fig:M123} .
    \label{fig:M12M13}}
  \end{figure}

\clearpage
 \begin{figure}[p!]
  \begin{center}
  \includegraphics[width=0.3\textwidth]{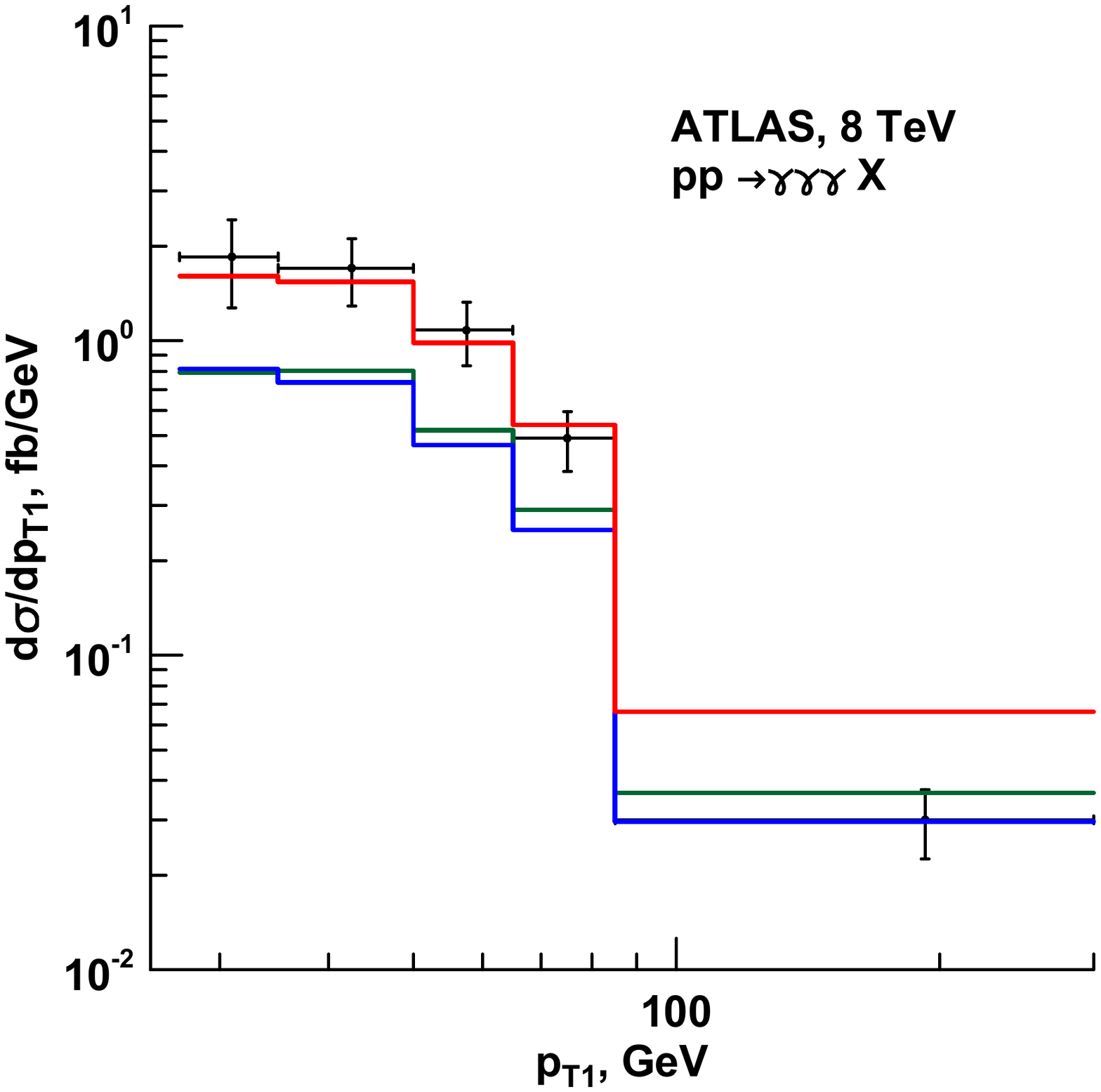}
  \includegraphics[width=0.3\textwidth]{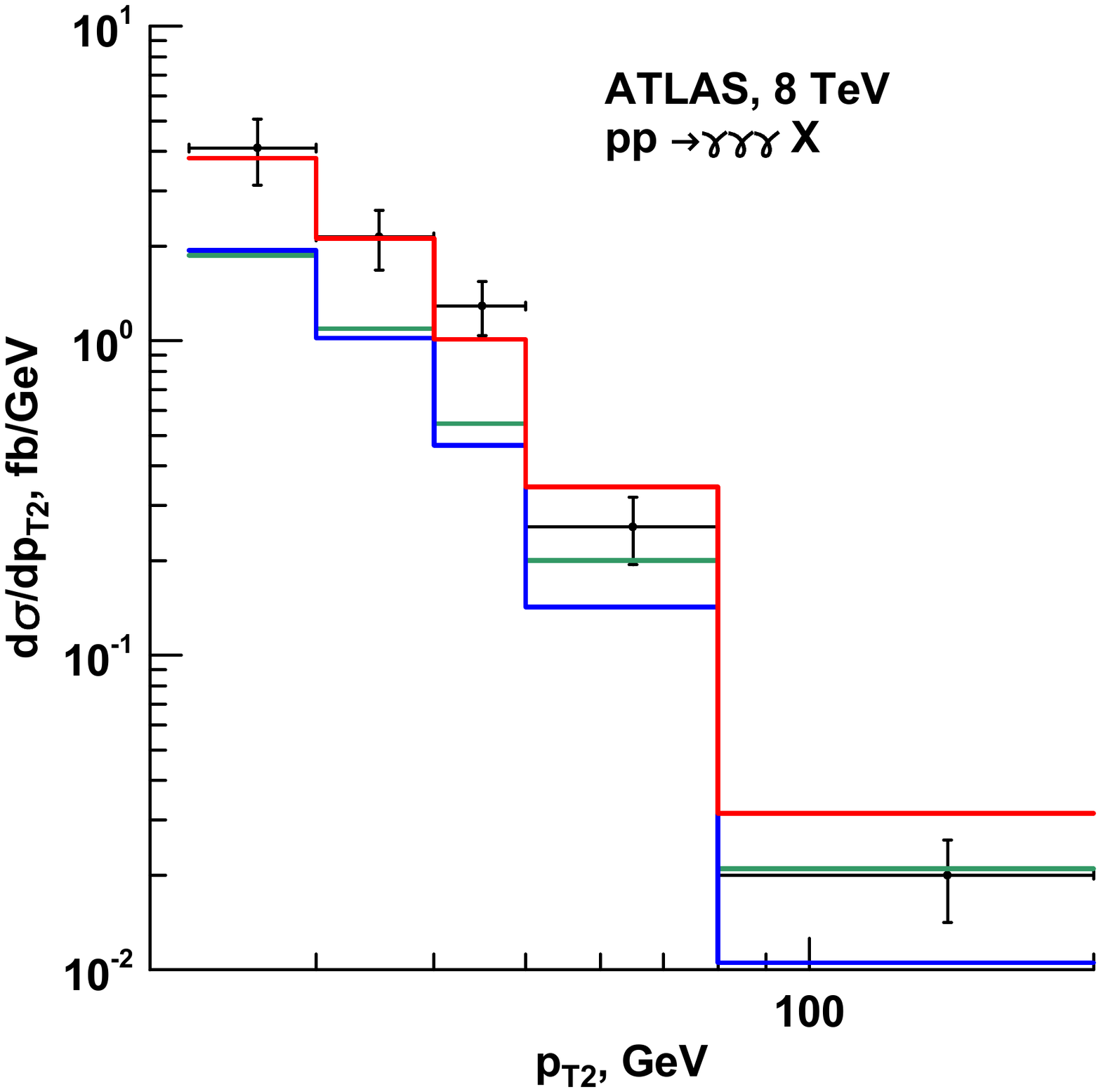}
  \includegraphics[width=0.3\textwidth]{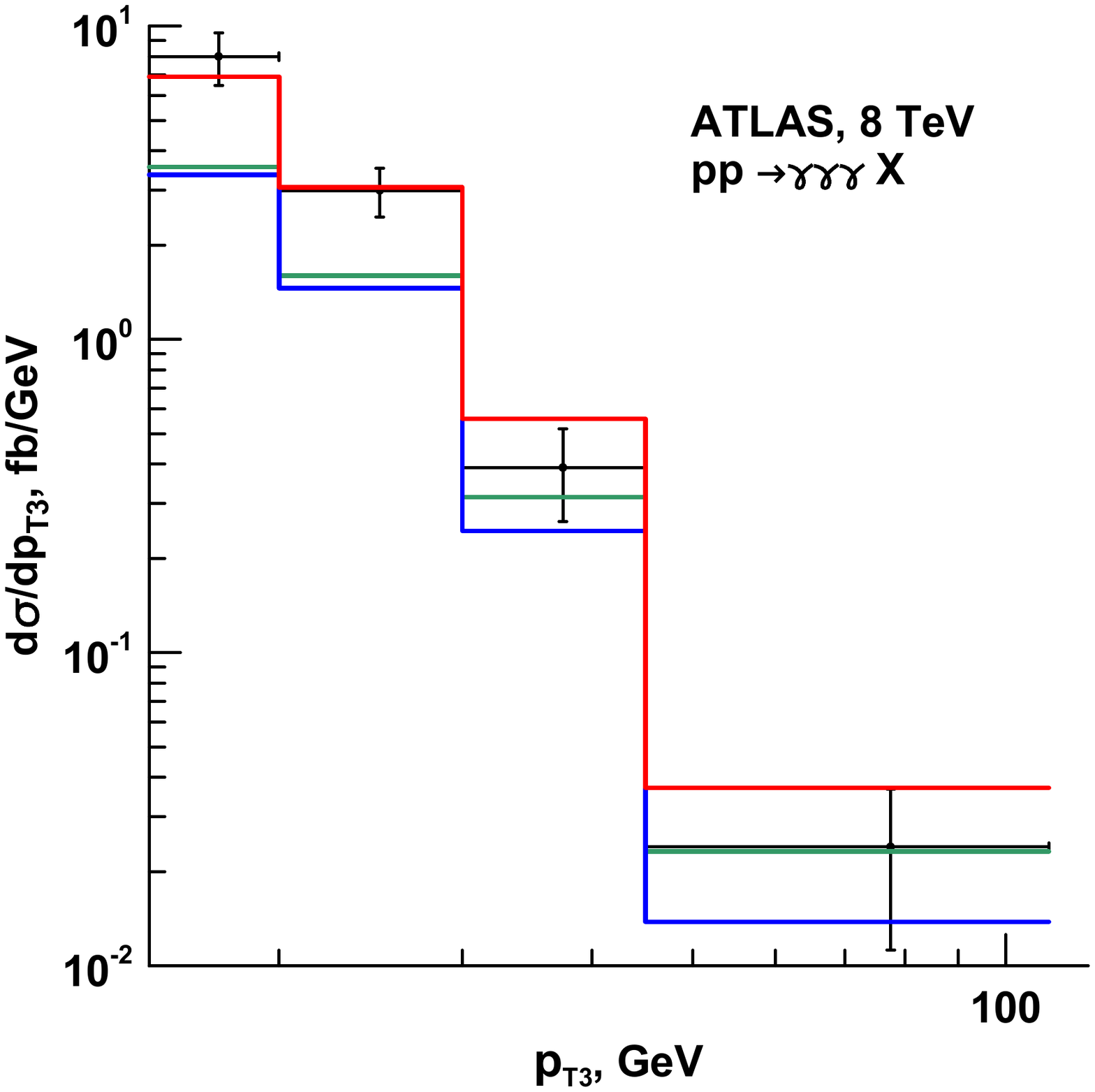}
  \end{center}
  \caption{The differential cross sections for the production of three isolated photons as functions
   of $p_{T1}$ (left panel), $p_{T2}$ (central panel) and $p_{T3}$ (right panel).
 Curves are defined as in Fig. \ref{fig:M123} .
    \label{fig:Pt12Pt13}}
  \end{figure}

\clearpage

\clearpage
 \begin{figure}[p!]
  \begin{center}
  \includegraphics[width=0.3\textwidth]{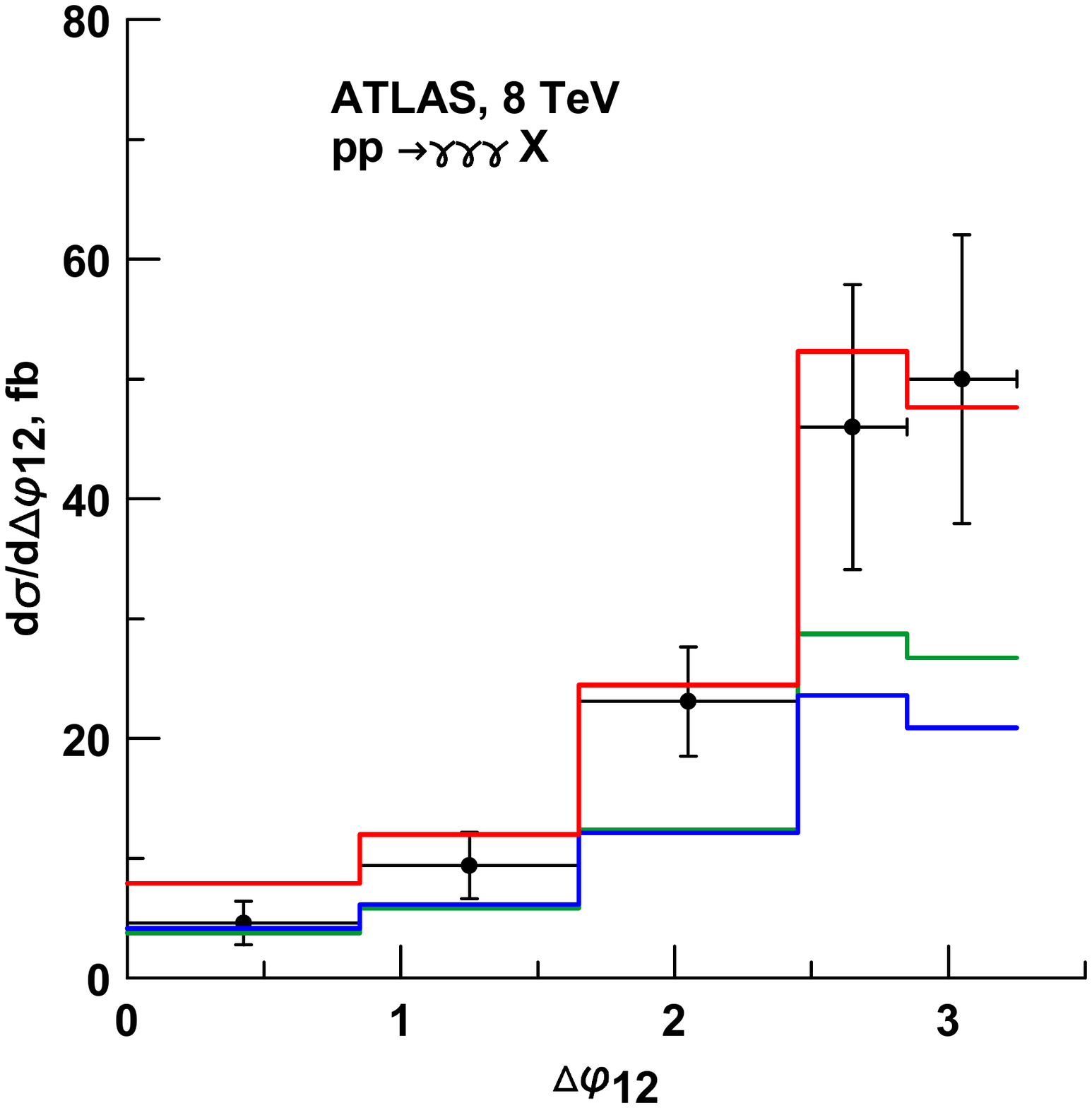}
  \includegraphics[width=0.3\textwidth]{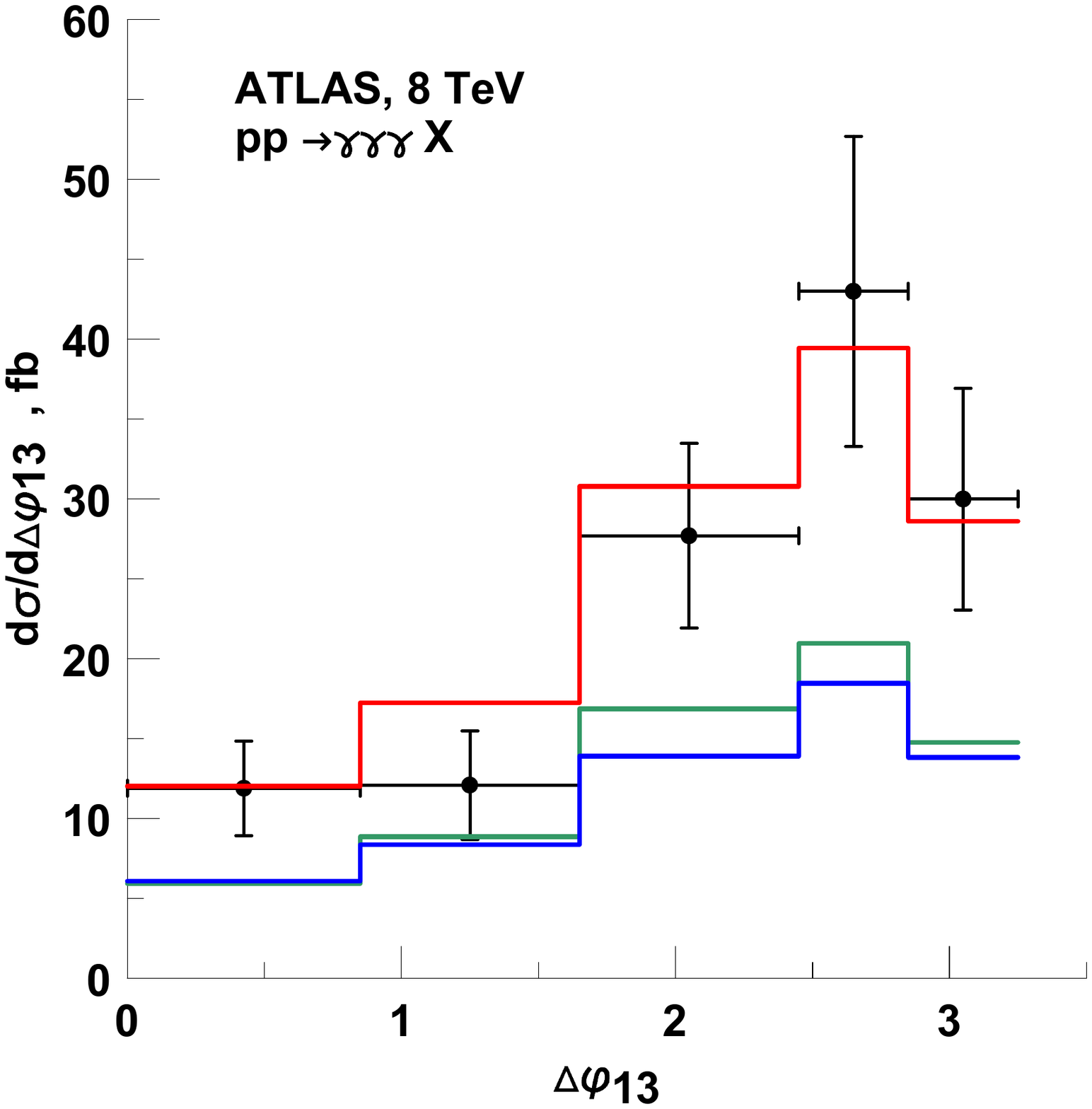}
  \includegraphics[width=0.3\textwidth]{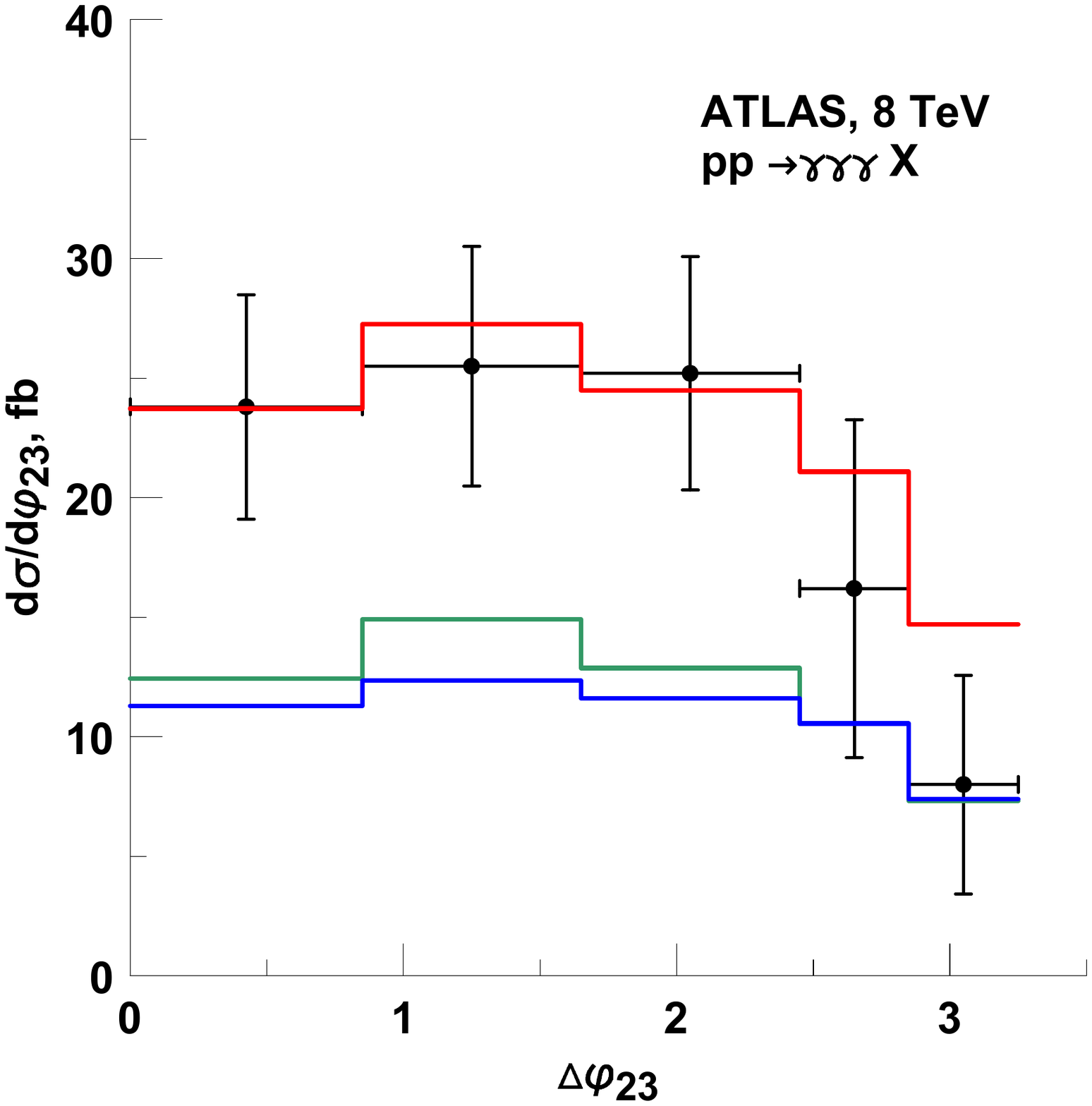}
  \end{center}
  \caption{The differential cross sections for the production of three isolated photons as functions
   of $|\Delta\phi_{12}|$ (left panel), $|\Delta\phi_{13}|$ (central panel) and $\Delta\phi_{23}$ (right panel).
   Curves are defined as in Fig. \ref{fig:M123} .
    \label{fig:Phi12Phi13}}
  \end{figure}

\clearpage

\clearpage
 \begin{figure}[p!]
  \begin{center}
  \includegraphics[width=0.3\textwidth]{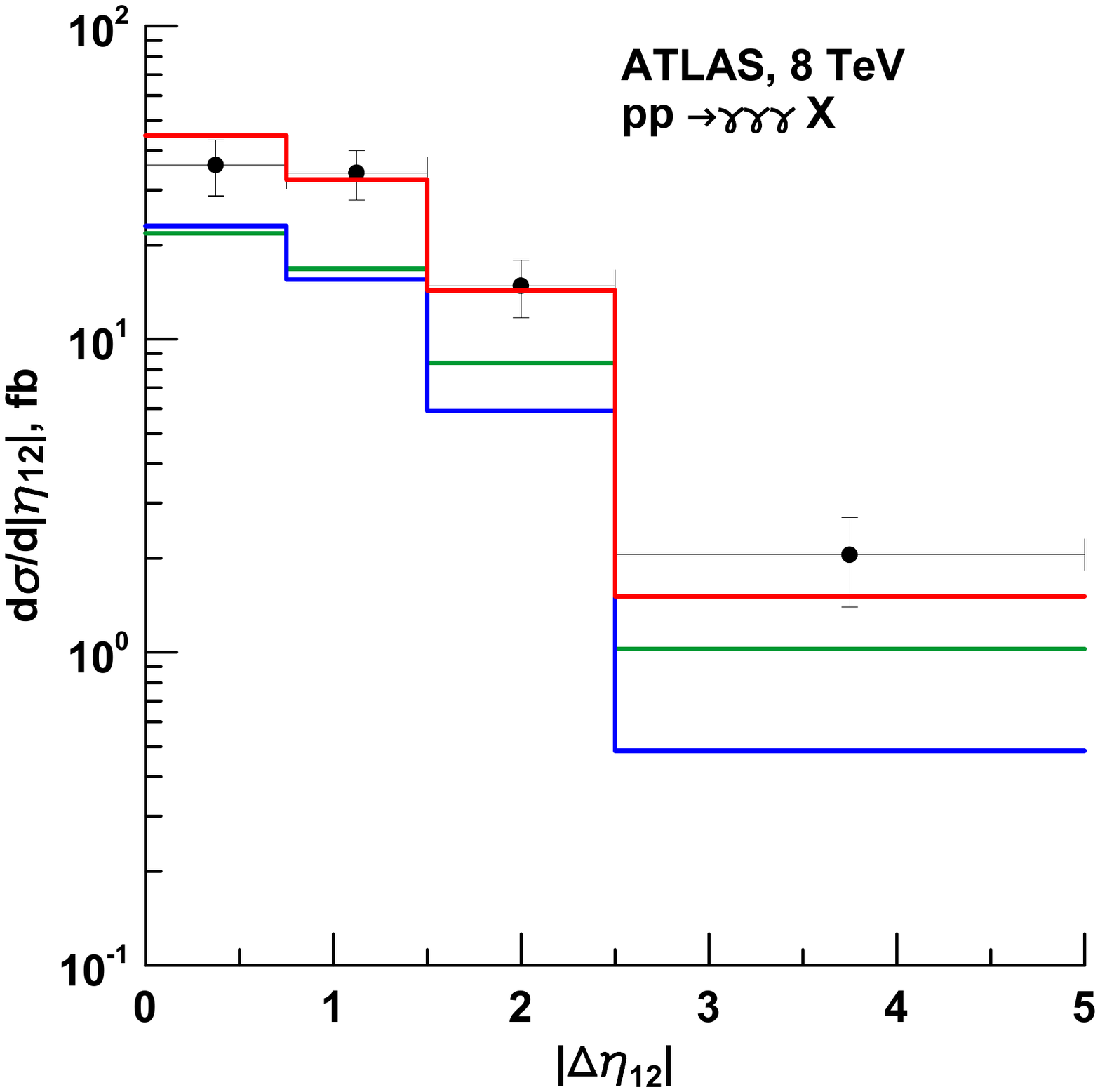}
  \includegraphics[width=0.3\textwidth]{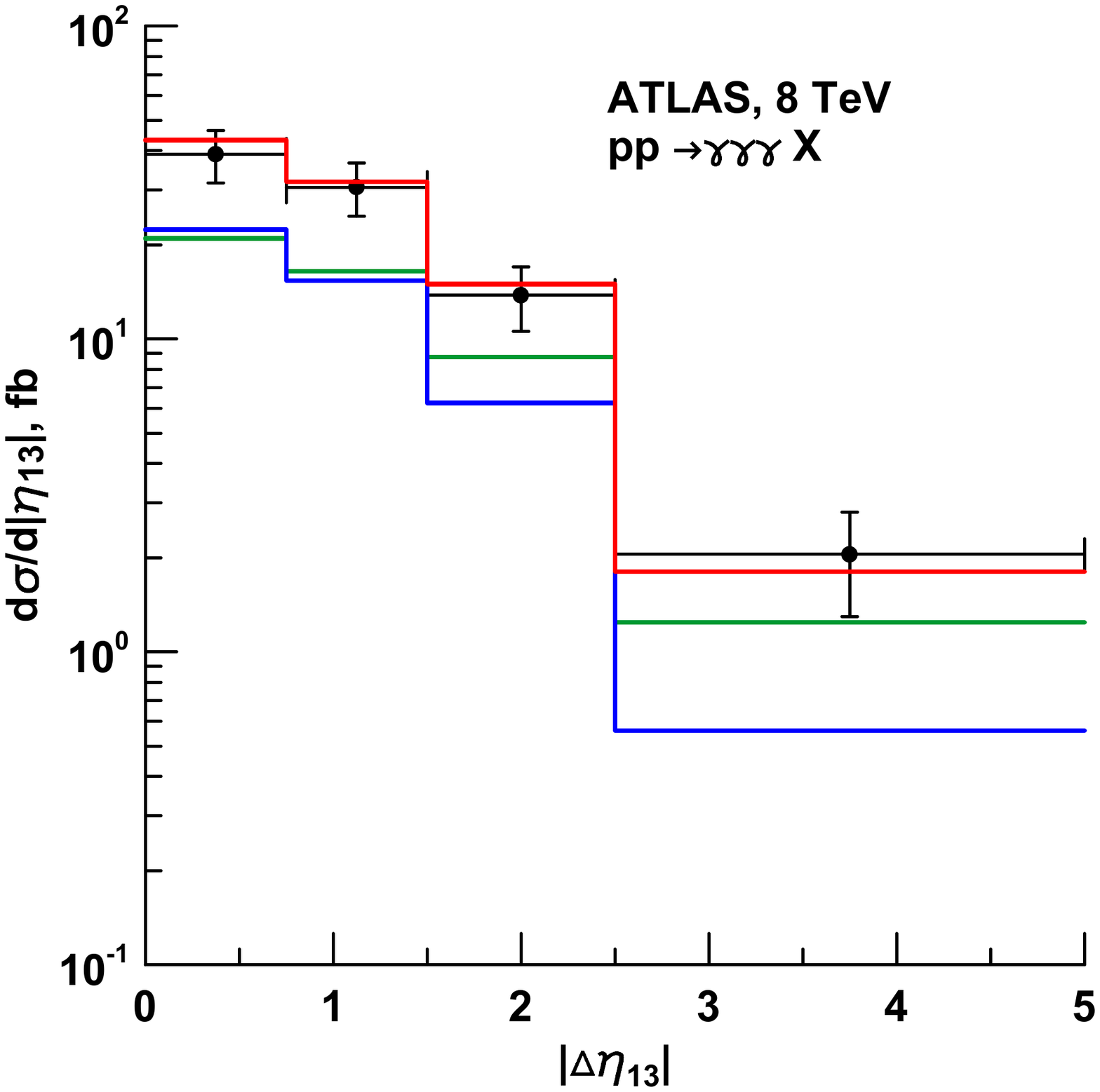}
    \includegraphics[width=0.3\textwidth]{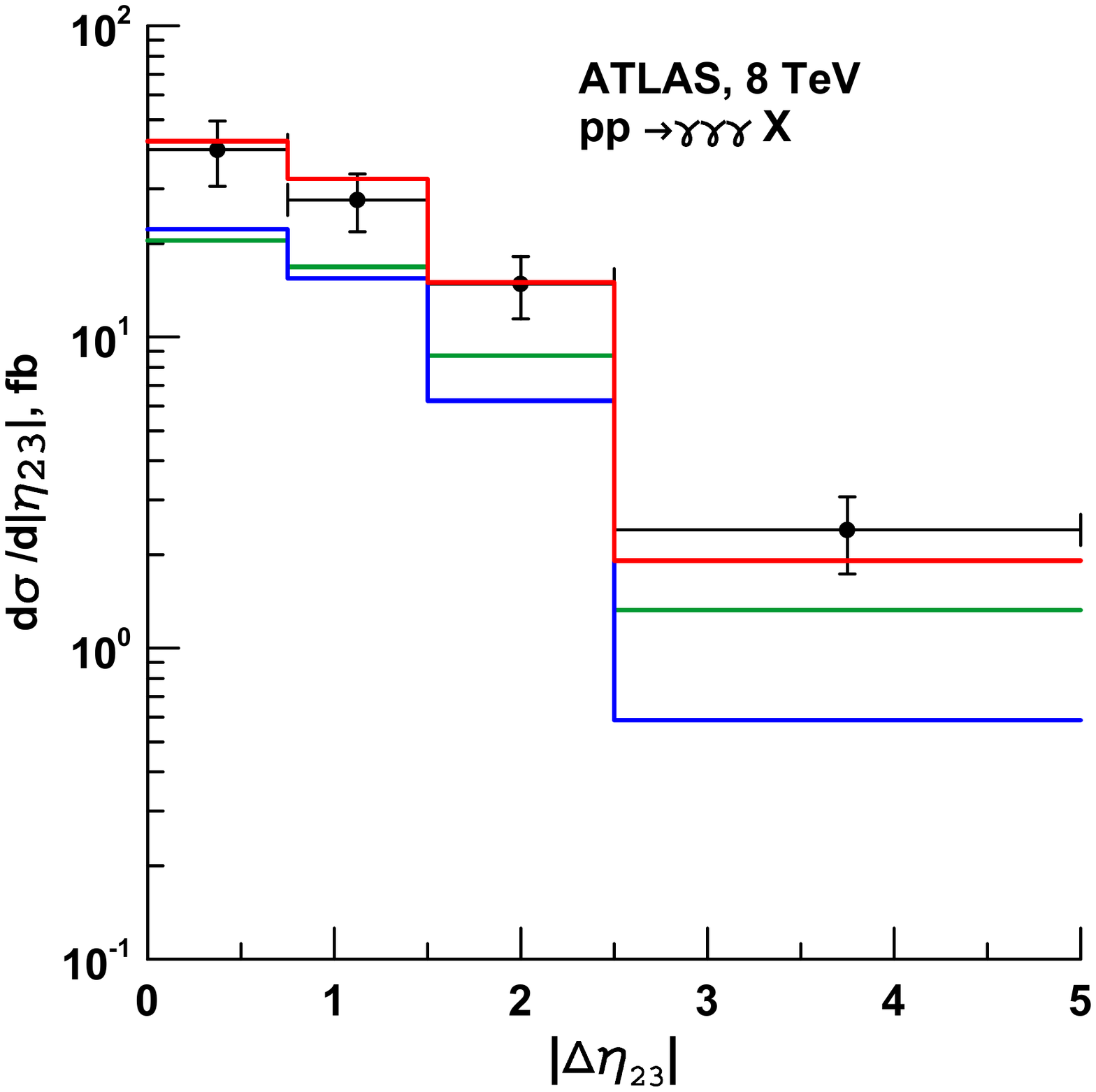}
  \end{center}
  \caption{The differential cross sections for the production of three isolated photons as functions
   of $|\Delta \eta_{12}|$ (upper left panel), $|\Delta \eta_{13}|$ (upper right panel) and $|\Delta \eta_{23}|$ (bottom panel).
   Curves are defined as in Fig. \ref{fig:M123} .
    \label{fig:Y12Y13}}
  \end{figure}

\end{document}